\documentclass[a4paper,11pt]{article}
\usepackage{pos}

\usepackage{graphicx}

\usepackage{subcaption}
\usepackage{svg}
\usepackage{amsmath}
\usepackage{booktabs}
\usepackage{multirow}
\usepackage{tikz}

\newlength{\bibitemsep}
\newlength{\bibparskip}
\let\oldthebibliography\thebibliography
\renewcommand\thebibliography[1]{%
  \oldthebibliography{#1}%
  \setlength{\parskip}{\bibitemsep}%
  \setlength{\itemsep}{\bibparskip}%
}

\title{Using Convolutional Neural Networks for the Helicity Classification of Magnetic Fields}

\ShortTitle{Using  Neural Networks for Helicity Classification}

\author*[a,b]{Nicol\`o Oreste Pinciroli Vago}
\author[a]{Ibrahim A.~Hameed}
\author[c]{M.~Kachelrie\ss}

\affiliation[a]{Institutt for IKT og realfag, NTNU, \AA lesund, Norway}

\affiliation[b]{Dipartimento di Elettronica, Informazione e Bioingegneria,
                Politecnico di Milano, Milan, Italy}

\affiliation[c]{Institutt for fysikk, NTNU, Trondheim, Norway}

\abstract{The presence of non-zero helicity in intergalactic magnetic fields is a smoking gun for their primordial origin since they have to be generated by processes that break CP invariance. As an experimental signature for the presence of helical magnetic fields, an estimator $Q$ based on the triple scalar product of the wave-vectors of photons generated in electromagnetic cascades from, e.g., TeV blazars, has been suggested previously. We propose to apply deep learning to helicity classification employing Convolutional Neural Networks and show that this method outperforms the $Q$ estimator.}

\FullConference{37th International Cosmic Ray Conference -ICRC2021-\\
                15-22 July 2021\\
                Online -- Berlin, Germany}

\begin{document}
\maketitle

\section{Introduction}

Magnetic fields are known to play a prominent role in the dynamics and the energy budget of astrophysical systems on galactic and smaller scales, but their role on larger scales is still elusive~\cite{Durrer:2013pga}. 
In galaxies and galaxy clusters, the observed magnetic fields are assumed to result from the amplification of much weaker seed fields. Such seeds could be created in the early universe, e.g.\ during phase transitions or inflation, and then amplified by plasma processes. If the generation mechanism of such primordial fields (e.g.\ by sphaleron processes) breaks CP, then the field will have a non-zero helicity. Since helical fields decay slower than non-helical ones, a small non-zero initial helicity is increasing with time, making the intergalactic magnetic field (IGMF) either completely left- or
right-helical today. A clean signature for a primordial origin of the IGMF is therefore its non-zero helicity. In a series of works, Vachaspati and collaborators worked out possible observational consequences of a helical IGMF, introducing the ``$Q$ statistics'' as a statistical estimator for the presence of helicity in the IGMF~\cite{Tashiro:2014gfa}.

In this work, we re-analyze the detection prospects of non-zero helicity in the IGMF using individual sources as, e.g., TeV blazars.
We propose the use of Deep Learning, a subfield of Machine Learning based on deep networks (i.e., networks with more than one layer), which can perform complex transformations on the original raw data in order to classify helicity data. A family of deep learning networks, in particular, is the one of Convolutional Neural Networks (CNN). They can be used for the analysis of tensor-encoded data as, e.g., images. Such networks have proven to be useful in many fields which are related to visual properties of images, ranging from artworks classification~\cite{milani2020data} to writer identification~\cite{8620507}, medicine~\cite{DBLP:journals/cbm/DeepakA19} and particle
physics~\cite{aurisano2016convolutional}.
Here, we evaluate the efficiency of different CNN models applied to synthetical
images of TeV blazars and compare the results to
a previous work using the $Q$ statistics~\cite{PhysRevD.102.083001}.
We describe the data transformation necessary to use
CNNs, present a quantitative comparison of the adopted CNN models, and
describe MobileNet~v2~\cite{MobileNetv2}, the most promising model among
those we investigated.

\section{Methodology}
\label{sec:methodology}

\begin{figure}[t]
    \centering
    \includegraphics[width=\textwidth]{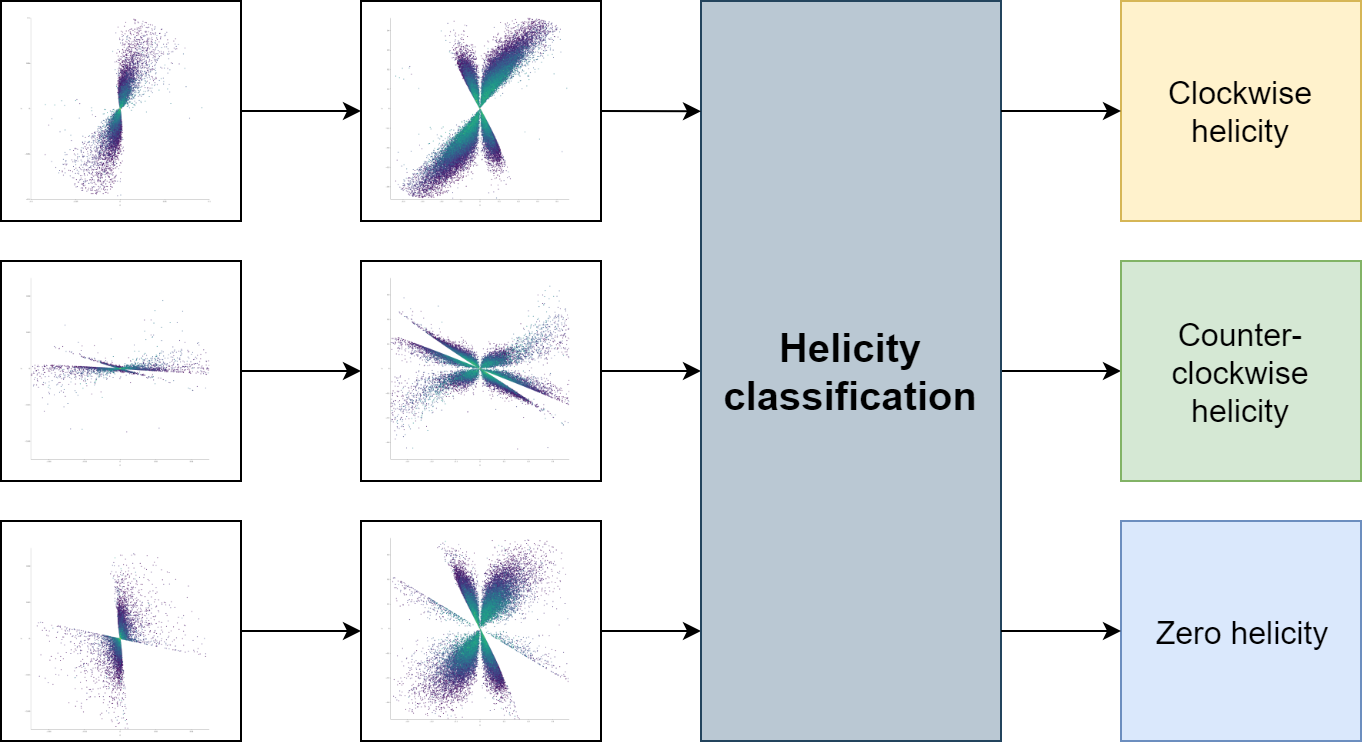}
 \caption{\textbf{The classifier as a black box.} The raw data are pre-processed and transformed in a more convenient format. The data processed in this way serve as input to the helicity classifier which in turn outputs the predicted helicity.}
    \label{fig:reduced_workflow}
\end{figure}

This section presents the classification workflow, from the generation of
mock data sets using {\tt ELMAG}~\cite{kachelriess2012elmag,blytt2020elmag}
to the application of a CNN to detect prominent visual patterns in the data.
Our approach consists of six main stages (see also Fig.~\ref{fig:workflow}
in the appendix):
\begin{enumerate}
  \setlength\itemsep{-0.5em}
    \item Generation of datasets using {\tt ELMAG}
    \item Data pre-processing
    \item Image generation
    \item Creation of the training, validation and test set
    \item Training and validation using a Convolutional Neural Network
    \item Testing of the obtained model on new data
\end{enumerate}
It is possible to consider the proposed classifier as a black box that, given an input image, outputs the corresponding helicity, as shown in Fig.~\ref{fig:reduced_workflow}.
This workflow can be applied for any possible initial simulation setting. In particular, it is possible to modify the data pre-processing parameters to give more importance to some parts of the plot (e.g., the central part), and to compare the performances and the metrics of different CNNs.

\subsection{Convolutional Neural Networks}

\begin{figure}[t]
  {\centering
    \vspace*{-1.5cm}
    \includegraphics{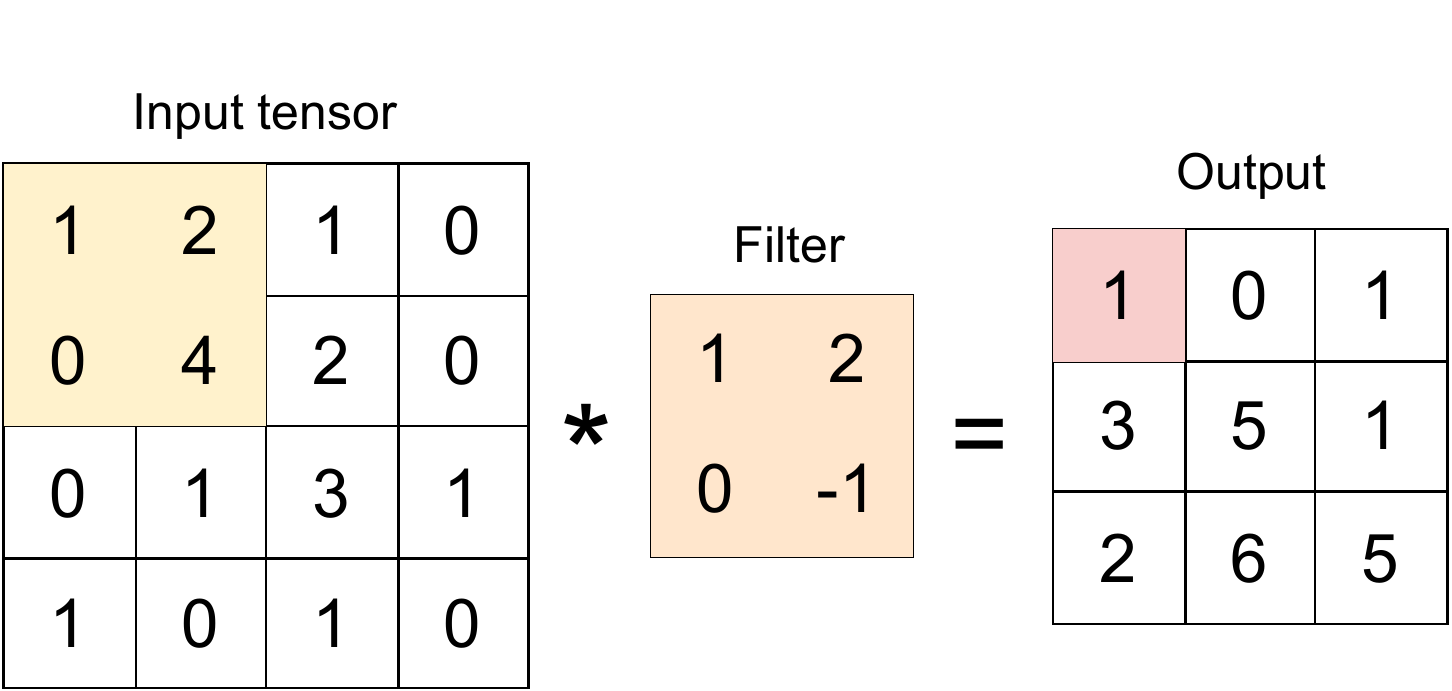}}
    \caption{A convolution with a single filter.
    \label{fig:convolution}}
\end{figure}

A CNN is a classifier that given an input datum represented as a tensor and a set of classes can predict the class (or, more in general, the classes) to which the datum belongs.
In our case, the input datum is an image of photons whose energy is colour
encoded and the output is the corresponding helicity label, $h=\{0,\pm 1\}$,
as presented in Fig.~\ref{fig:reduced_workflow}.
More precisely, CNNs are a class of artificial neural networks (ANN), which, differently from traditional ANNs, can perform convolutions in one or more dimensions using convolutional layers. In particular, convolutions are defined by introducing one or more filters, which are tensors of numbers.
A filter is originally placed in correspondence of the top-left corner of the tensor representing an input datum (e.g.\ an image), and an element-wise multiplication between the filter and the underlying datum elements is performed. The multiplied elements are then summed and placed in an output tensor, in the same position as the filter top-left corner position. The filter, then, is moved by a quantity called \textit{stride} along all the tensor elements, until the entire tensor has been covered. It is also possible to introduce \textit{padding}, which consists of adding zeros to the tensor borders to obtain an output tensor with the same dimensions as the input tensor. Figure~\ref{fig:convolution} presents the result of a convolution operation on a bidimensional tensor with a single filter.
The highlighted elements represent the first operation performed by the
convolution, which in this example is given by
      $\big(\begin{smallmatrix}
1 & 2\\
0 & 4
\end{smallmatrix}\big) * \big(\begin{smallmatrix}
1 & 2\\
0 & -1
\end{smallmatrix}\big) = 1\cdot 1 + 2\cdot 2 + 0\cdot 0 + 4\cdot (-1) =1$.

Additional to the convolution, many CNNs include pooling operations which
output, for each position of the filter, the maximum element below the
filter (\textit{max pooling\/}) or the average of the elements below the filter
(\textit{average pooling\/}).
In the final part of a CNN, it is necessary to insert fully connected layers, which consider all the inputs from the previous layer, perform a linear combination of such inputs, and output a vector whose length corresponds to the number of classes of the problem (i.e., the helicity cases). The content of this vector consists of the probabilities associated with the different classes. A CNN, then, is formed by a sequence of convolutional layers, pooling layers, more complex layers based on them, and fully connected layers. This means that it encodes a complex transformation of the initial data
into the labels associated with them.
When images are considered, a convolution is defined over three dimensions (i.e., the width, the height and the depth). The depth initially allows encoding the colour channels in the RGB format, arranging them in subsequent layers. During the learning process, each layer of the transformed tensor represents a filter. Applying a filter over a tensor (e.g., the input image or its transformations) further transforms it. Consequently, the input tensor has the shape $n \times m \times 3$ for a coloured image with $n \times m$ pixels. After the first convolution, the output tensor will have the shape $n' \times m' \times f$ where $f$ is the number of filters, and $n'$ and $m'$ depend on the filter size and the presence of padding.

\subsection{Dataset generation}

The first stage of the workflow consists of the generation of an initially labelled dataset using {\tt ELMAG}. In particular, in this first stage, the dataset contains the sky positions and energies of the simulated photons in a tabular form and the corresponding label (i.e., for each simulation, the helicity value of the turbulent field, as set in {\tt ELMAG}). For each of the three helicity cases, 1160~photons reaching the observer were simulated.
It is in principle possible to generate a larger dataset,
but as the dataset size increases, the time required for training and the storage requirements increase.

\subsection{Data pre-processing and image generation}

Data pre-processing is essential to transform the initial raw tabular
data in a more convenient format, which can be used for the generation of plots formatted as images and used as the input of CNN models. 
Energy is not encoded automatically in a bi-dimensional plot.
Since energy is not automatically encoded in a bidimensional plot, we use
the colour channel to represent different energy levels, associating colour
to the logarithm of the photon energy. This continuous colour
encoding of energy has to be compared to the use of energy bins in the
calculation of the $Q$ estimator.

\begin{figure*}[t]
        \begin{subfigure}[b]{0.49\textwidth}
            \centering
            \includegraphics[width=\textwidth]{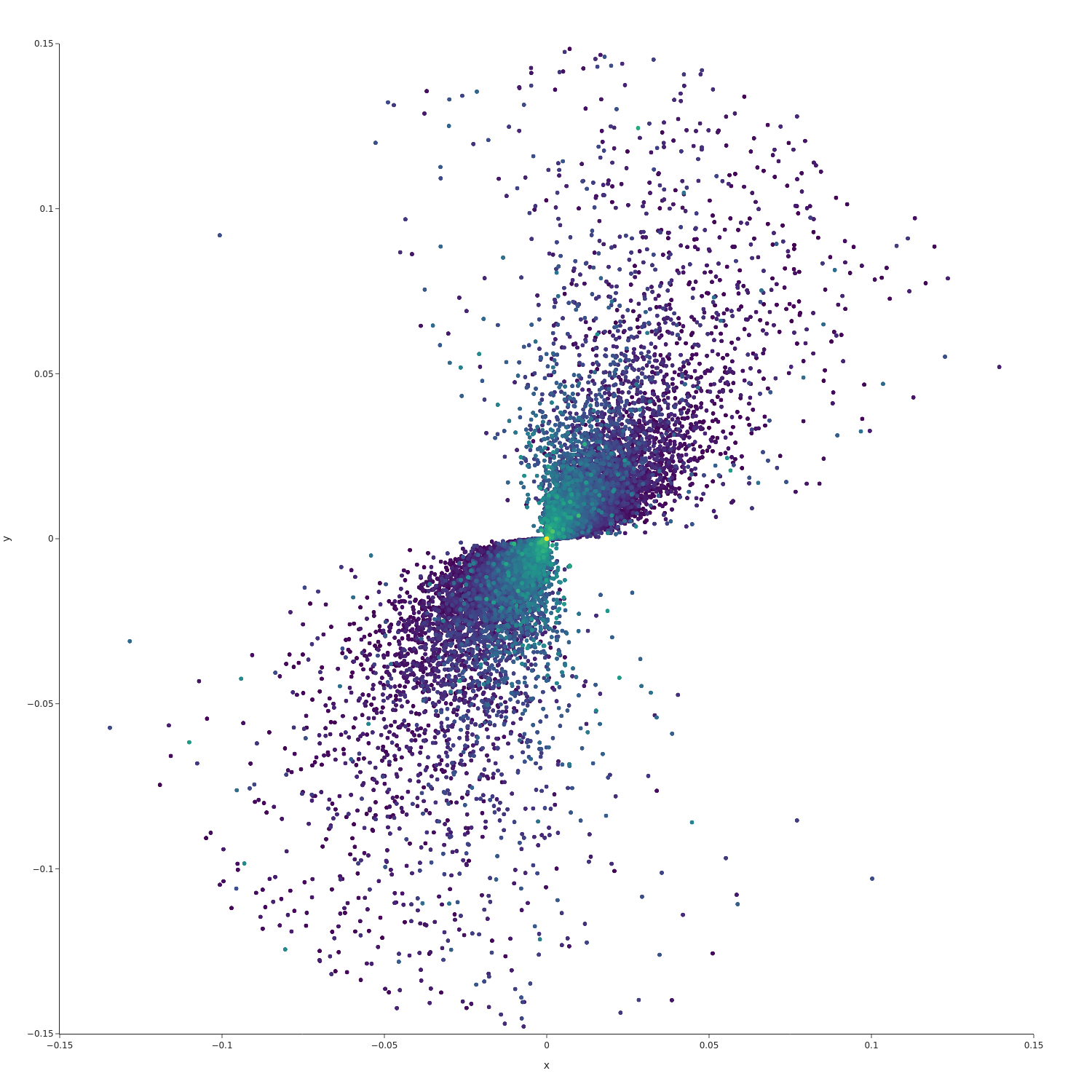}
            \caption{The original plot.}
        \end{subfigure}
\hspace{0em}
        \begin{subfigure}[b]{0.49\textwidth}
            \centering
            \includegraphics[width=\textwidth]{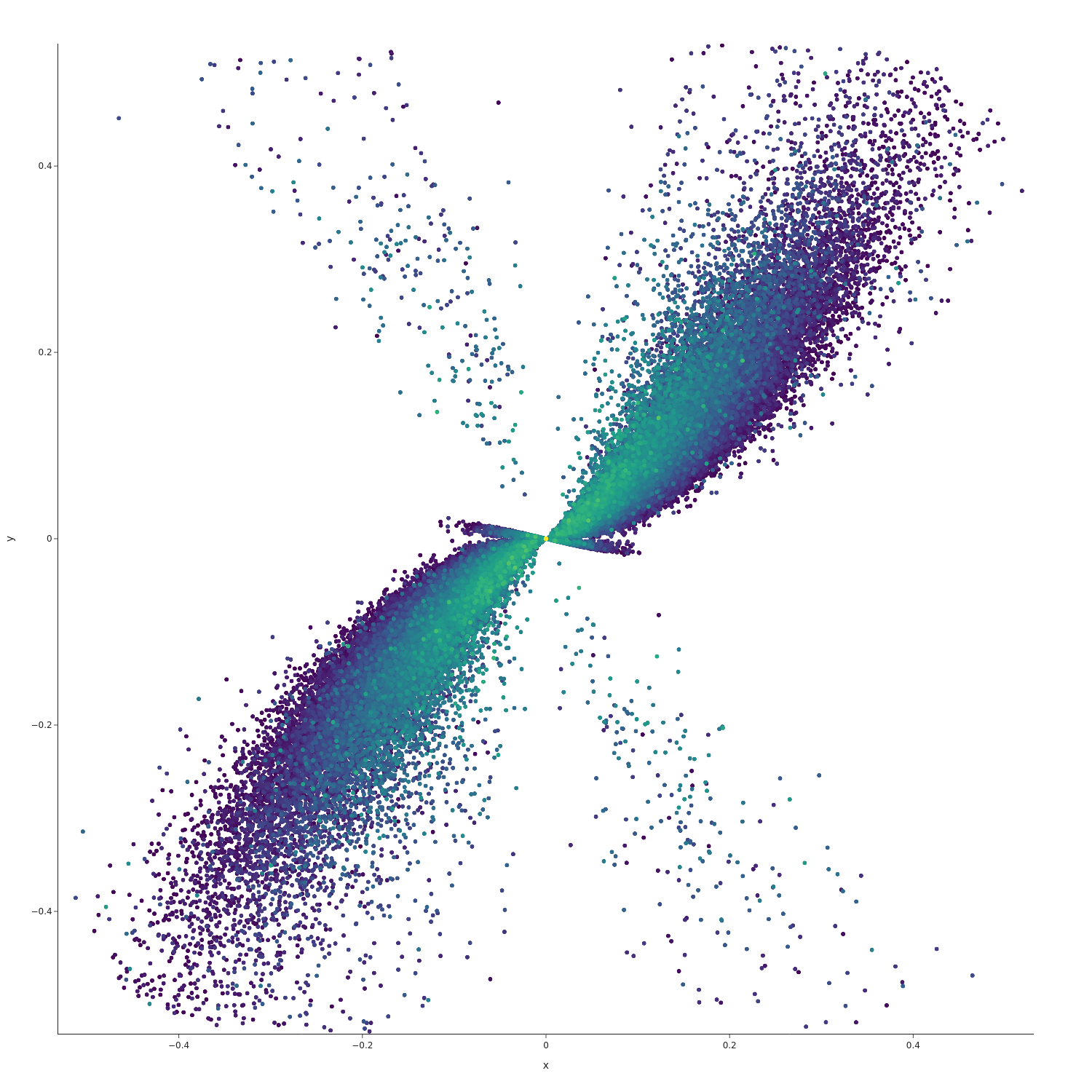}
            \caption{The plot after the transformation.}
        \end{subfigure}

        \caption{\textbf{Image transformation.} Left: Original image for a counter-clockwise helix. Right: After the cubic root operation. Energy  is encoded as colour, varying from  yellow (highest energies) to blue (lowest).}
        \label{fig:transformation}
    \end{figure*}

We also encode the positions differently than in the original data,
to visually emphasize the helicity direction. In the first step, we
remove the 1\% of photons with the largest distance to the origin. Then we introduce dimensionless $x,y$ coordinates and, if necessary, rescale them such that the coordinates of all remaining photons are contained in the range $[-1, +1]$. Next, we want to transform the image to emphasize its overall helical curvature. A set of functions leaving the sign of the $x,y$ coordinates unchanged is defined, for a single coordinate, by $F = \{x^n \,|\, n=\frac{1}{2k+1} \wedge k\in\mathbb{N} \wedge x\in\mathbb{R}\}$.
In particular, the cubic root is part of this set, and we chose it to transform the initial data since it is the lowest order transformation in the set. Figure~\ref{fig:transformation} shows an example of this transformation. While the original image looks almost
symmetrical with respect to the bisector of the first and third quadrant, the image after the transformation is less symmetrical and emphasizes the counter-clockwise helicity. Finally, the image dimensions used as the network input are set to 500 $\times$ 500 pixels.

\subsection{Data splitting and training}

Once the data are generated, they are sorted into a training (70\% of the total data), validation (15\%) and test set (15\%). The training data are used for the training of the CNN (i.e., during the network learning phase). The validation data are used to validate the results from the training to check whether there is overfitting (i.e., whether the network memorizes the data, but is not able to generalize on new data). The test set is used, after the training and the validation, on data previously unknown to the network. In this work, we generate also the test data using {\tt ELMAG}. However, the same procedure can be applied to real data, if they are pre-processed in the same way as the simulated data.

\subsection{Architecture}

The proposed classifier exploits the MobileNet~v2
architecture~\cite{MobileNetv2}, pre-trained on the ImageNet~\cite{5206848}
data set. We decided to use a pre-trained model, rather than performing the
entire training from scratch, to exploit the initial layers' content,
which is expected to contain filters for the recognition of simple patterns and geometrical shapes. Moreover, this choice allowed us to use fewer images in the training process, which corresponds to a shorter overall computational time spent on the initial simulation. Since the images' complexity and variability are rather low across our data set, we did not perform data augmentation.

MobileNet~v2's main building block, differently from traditional CNNs, is a residual block that contains three convolutional layers: the expansion layer, the depth-wise convolutional layer and the projection layer. The expansion layer has the purpose of expanding the number of channels in the data. Given a $n \times m \times f$ tensor, its purpose is to generate a $n \times m \times cf$ tensor, where $c \in \mathbb{N}$ is defined as the expansion factor. To perform this operation, the expansion layer applies a $1\times 1$ convolution considering $cf$ filters. The depth-wise convolutional layer, differently from the traditional convolutional layer,
considers each feature map separately, i.e. the input tensor feature maps are not combined. The final layer of this block is the projection layer, which reduces the number of feature maps by projecting high-dimensional data into a lower-dimensional tensor. Finally, the CNN contains residual blocks, which aim at propagating promising results in complex networks. The residual blocks introduce, in addition to the sequential path starting from the input tensor and ending in the prediction result, connections that can skip one or more blocks. The outputs of such connections are summed to the outputs obtained in the main path.

\section{Results and Discussion}
\label{sec:results}

\begin{figure*}[t]
        \begin{subfigure}[b]{0.49\textwidth}
            \centering
            \includegraphics[width=\textwidth]{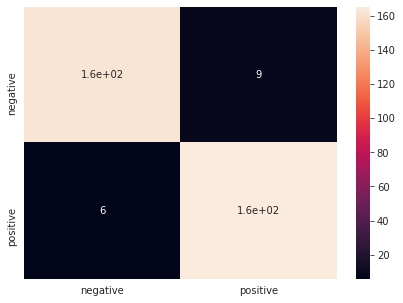}
            \caption{Clockwise and counter-clockwise helicity}
            \label{fig:confusion2}
        \end{subfigure}
\hspace{0em}
        \begin{subfigure}[b]{0.49\textwidth}
            \centering
            \includegraphics[width=\textwidth]{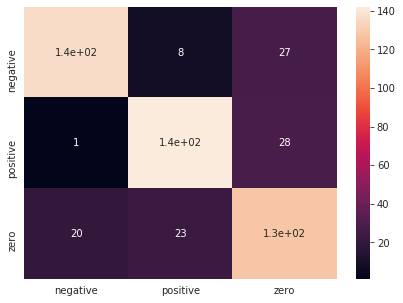}
            \caption{Clockwise, counter-clockwise and zero helicity}
            \label{fig:confusion3}
        \end{subfigure}
        \caption{\textbf{Confusion matrices for clockwise (positive), counter-clockwise (negative) and zero helicity.} }
        \label{fig:confusion}
    \end{figure*}

\begin{figure*}[t]
        \begin{subfigure}[b]{0.49\textwidth}
            \centering
            \includegraphics[width=\textwidth]{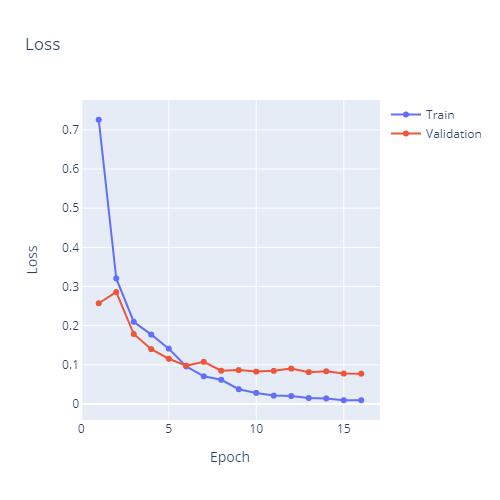}
            \caption{Train and validation loss}
            \label{fig:2classloss}
        \end{subfigure}
\hspace{0em}
        \begin{subfigure}[b]{0.49\textwidth}
            \centering
            \includegraphics[width=\textwidth]{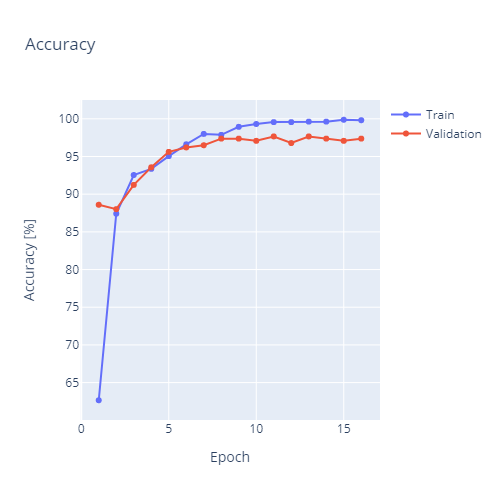}
            \caption{Train and validation accuracy}
            \label{fig:2classaccuracy}
        \end{subfigure}
        \caption{\textbf{Train and validation loss and accuracy for two classes.} Loss and accuracy for MobileNet v2, with a learning rate $\eta = 0.01$, a batch size of 4 and 16 epochs.}
        \label{fig:2classes_loss_acc}
    \end{figure*}

We choose physical parameters as in Ref.~\cite{PhysRevD.102.083001}, to compare to their results. In particular, we use as strength of the turbulent
magnetic field $B = 10^{-16}$\,~G, set 10\,GeV as low energy cut-off,
and neglect the finite angular resolution of realistic experiments. We
consider how well CNNs can distinguish positive and negative helicity, while results including zero helicity fields are discussed in the appendix. Table~\ref{tab:2classes} shows a comparison among the accuracy of several CNN models, which achieve an accuracy between 87\% and 96\%. MobileNet~v2 accomplishes the highest accuracy among all networks, followed by ShuffleNet. The
96\%~accuracy of  MobileNet~v2  corresponds to an overlap of 8\% in the language of Ref.~\cite{PhysRevD.102.083001},
surpassing boldly the 53\% overlap obtained with the $Q$ estimator.
Figure~\ref{fig:confusion2} represents the confusion matrix for  MobileNet~v2, while Fig.~\ref{fig:2classes_loss_acc} presents the validation loss during the training. The greatest confusion is observed for the zero helicity case since it is an intermediate case and is more easily confused with the clockwise (or positive) or counter-clockwise (or negative) helicity cases. Moreover, one can see that the loss decreases both for the training and for the validation set, which suggests that the network has good generalization abilities.

\begin{table}[]
\centering
\caption{\textbf{Comparison of models for the classification of clockwise and counter-clockwise helicity.} This table shows, for each model and each hyperparameter setting, the accuracy on the train set, the validation set and the test set. Here $\eta$ refers to the learning rate, while the batch size refers to the number of images considered simultaneously during the training phase. All the experiments used Adagrad as the optimizer and were based on networks pre-trained on ImageNet.}

\label{tab:2classes}
\resizebox{\textwidth}{!}{%
\begin{tabular}{@{}cccccccc@{}}
\toprule
\textbf{Model}                       & \textbf{$\eta$ }   & \textbf{$\eta$ decay} & \textbf{Batch size} & \textbf{\# epochs} & \textbf{Train accuracy} & \textbf{Validation accuracy} & \textbf{Test accuracy} \\ \midrule
ResNet50                    & 0,010 & 0,0008   & 2          & 15        & 93,80\%        & 94,74\%             & 91,00\%       \\ \midrule
ResNet18                    & 0,010 & 0,0008   & 2          & 12        & 95,43\%        & 95,32\%             & 93,00\%       \\ \midrule
GoogleNet                   & 0,010 & 0,0008   & 2          & 8         & 93,73\%        & 92,11\%             & 90,00\%       \\ \midrule
\multirow{3}{*}{MobileNet v2}  & 0,010 & 0,0008   & 2          & 10        & 97,62\%        & 94,44\%             & 95,00\%       \\ \cmidrule(l){2-8} 
                            & 0,001 & 0,0008   & 2          & 16        & 95,43\%        & 93,86\%             & 90,00\%       \\ \cmidrule(l){2-8} 
                            & 0,010 & 0,0008   & 4          & 16        & 99,81\%        & 97,37\%             & 96,00\%       \\ \midrule
\multirow{4}{*}{ShuffleNet} & 0,010 & 0,0008   & 2          & 8         & 99,12\%        & 94,15\%             & 95,00\%       \\ \cmidrule(l){2-8} 
                            & 0,005 & 0,0008   & 2          & 6         & 97,50\%        & 95,32\%             & 95,00\%       \\ \cmidrule(l){2-8} 
                            & 0,005 & 0,0008   & 8          & 6         & 99,38\%        & 95,03\%             & 95,00\%       \\ \cmidrule(l){2-8} 
                            & 0,005 & 0,0008   & 4          & 11        & 99,69\%        & 95,91\%             & 92,00\%       \\ \midrule
DenseNet121                 & 0,010 & 0,0008   & 2          & 10        & 90,29\%        & 91,52\%             & 87,00\%       \\ \midrule
HardNet68                   & 0,010 & 0,0008   & 2          & 17        & 90,16\%        & 89,47\%             & 87,00\%       \\ \midrule
GhostNet                    & 0,010 & 0,0008   & 2          & 14        & 99,69\%        & 94,44\%             & 92,00\%       \\ \bottomrule
\end{tabular}%
}
\end{table}

\section{Conclusions}
\label{sec:conclusions}

We have analyzed the application of Convolutional Neural Networks to the problem of predicting the helicity of the IGMF using synthetical images of TeV blazars. Among the networks examined, MobileNet~v2 achieved the highest accuracy and the results obtained outperform those derived using the $Q$ estimator. While the obtained increase in accuracy is encouraging, additional
research is needed to address more realistic cases. In particular,
the angular resolution and energy-dependent sensitivity of specific
experiments will deteriorate the results presented here for the case of an "ideal'' experimental set-up. Another question to address is if the
data pre-processing can be improved further, by changes in the colour-coding of energy or transformations alternative to the cubic root.



\newpage\clearpage
\appendix
\section{Appendix: Details}
\label{app}

\begin{figure}
    \centering
    \includegraphics[height=0.6\textheight]{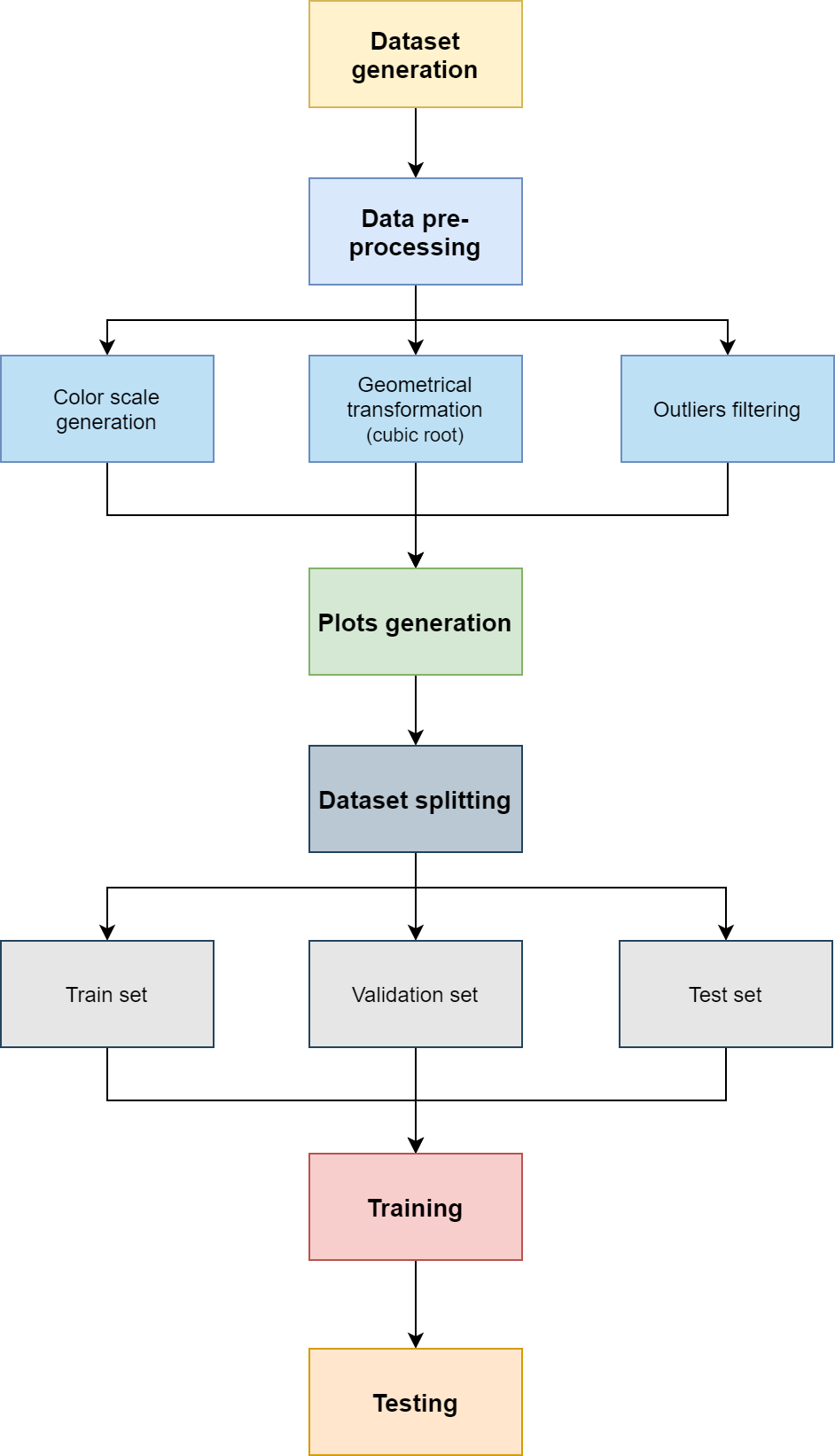}
    \caption{\textbf{The overall workflow.} This diagram presents the stages performed for the development of a helicity classifier. In particular, the main steps consist of the automatic generation of the dataset, the data transformation to ease the subsequent classification, the actual training and the testing on new automatically generated data, previously unknown to the classifier.}
    \label{fig:workflow}
\end{figure}

Figure~\ref{fig:workflow} illustrates the classification workflow, from the generation of mock data to the application of the CNN. In Table~2, we compare the classification accuracy achieved including the zero helicity case for three CNNs.

\begin{table}[]
\centering
\caption{\textbf{Comparison of models for the classification of zero, clockwise and counter-clockwise helicity.} This table shows, for each model and each hyperparameter setting, the accuracy on the train set, on the validation set and on the test set. Here $\eta$ refers to the learning rate and the batch size refers to the number of images considered simultaneously during the training phase. All the experiments used Adagrad as the optimizer, and were based on networks pre-trained on ImageNet.}
\label{tab:3classes}
\resizebox{\textwidth}{!}{%
\begin{tabular}{@{}cccccccc@{}}
\toprule
\textbf{Model}              & \textbf{$\eta$} & \textbf{$\eta$ decay} & \textbf{Batch size} & \textbf{\# epochs} & \textbf{Train accuracy} & \textbf{Validation accuracy} & \textbf{Test accuracy} \\ \midrule
ResNet18                    & 0,005       & 0,0010            & 2                   & 6                  & 78,40\%                 & 77,58\%                      & 77,00\%                \\ \midrule
\multirow{3}{*}{MobileNet v2}  & 0,005       & 0,0010            & 2                   & 8                  & 84,34\%                 & 79,73\%                      & 79,00\%                \\ \cmidrule(l){2-8} 
                            & 0,010       & 0,0008            & 4                   & 6                  & 79,76\%                 & 79,73\%                      & 78,00\%                \\ \cmidrule(l){2-8} 
                            & 0,005       & 0,0010            & 4                   & 5                  & 80,30\%                 & 80,90\%                      & 79,00\%                \\ \midrule
\multirow{4}{*}{ShuffleNet} & 0,005       & 0,0010            & 4                   & 6                  & 88,89\%                 & 80,70\%                      & 78,00\%                \\ \cmidrule(l){2-8} 
                            & 0,005       & 0,0010            & 2                   & 6                  & 82,96\%                 & 78,17\%                      & 75,00\%                \\ \cmidrule(l){2-8} 
                            & 0,010       & 0,0008            & 2                   & 6                  & 83,38\%                 & 76,02\%                      & 74,00\%                \\ \cmidrule(l){2-8} 
                            & 0,005       & 0,0010            & 8                   & 9                  & 87,51\%                 & 77,39\%                      & 75,00\%                \\ \bottomrule
\end{tabular}%
}
\end{table}

%
%

        \begin{figure*}[t]
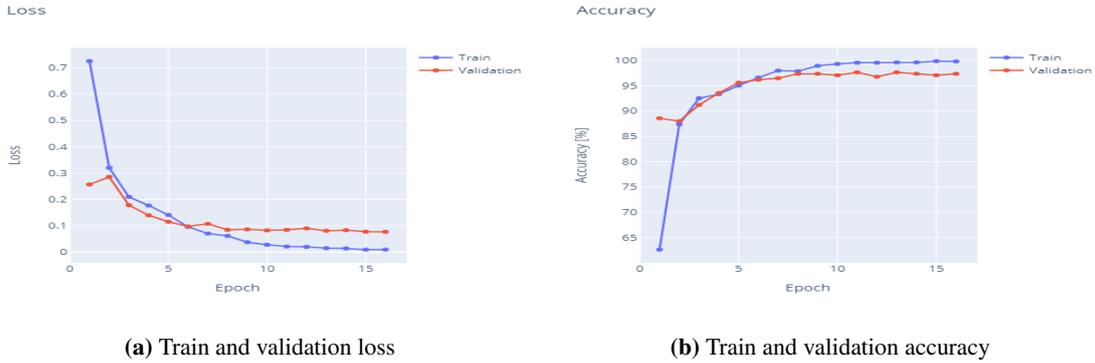

        \begin{subfigure}[b]{0.49\textwidth}
            \centering
            \includegraphics[width=\textwidth,height=0.6\textwidth]{images/2_classes_loss.png}
            \caption{Train and validation loss}
            \label{fig:3classloss}
        \end{subfigure}
\hspace{0em}
        \begin{subfigure}[b]{0.49\textwidth}
            \centering
            \includegraphics[width=\textwidth,height=0.6\textwidth]{images/2_classes_accuracy.png}
            \caption{Train and validation accuracy}
            \label{fig:3classaccuracy}
        \end{subfigure}

        \caption{\textbf{Train and validation loss and accuracy for three classes.} Those plots show the results in terms of loss and accuracy for MobileNet v2, with a learning rate $\eta = 0.005$, a batch size of 4 and 5 epochs. It is possible to notice that the loss decreases both for the train set and for the validation set, which suggests that the network has good generalization abilities.}
        \label{fig:3classes_loss_acc}
    \end{figure*}

\subsection{Case studies}

This section presents some case studies with both correct and incorrect predictions for the three classes under analysis.

\subsubsection{Clockwise helicity}

\begin{figure*}[!ht]
        \begin{subfigure}[b]{0.49\textwidth}
            \centering
            \includegraphics[width=\textwidth]{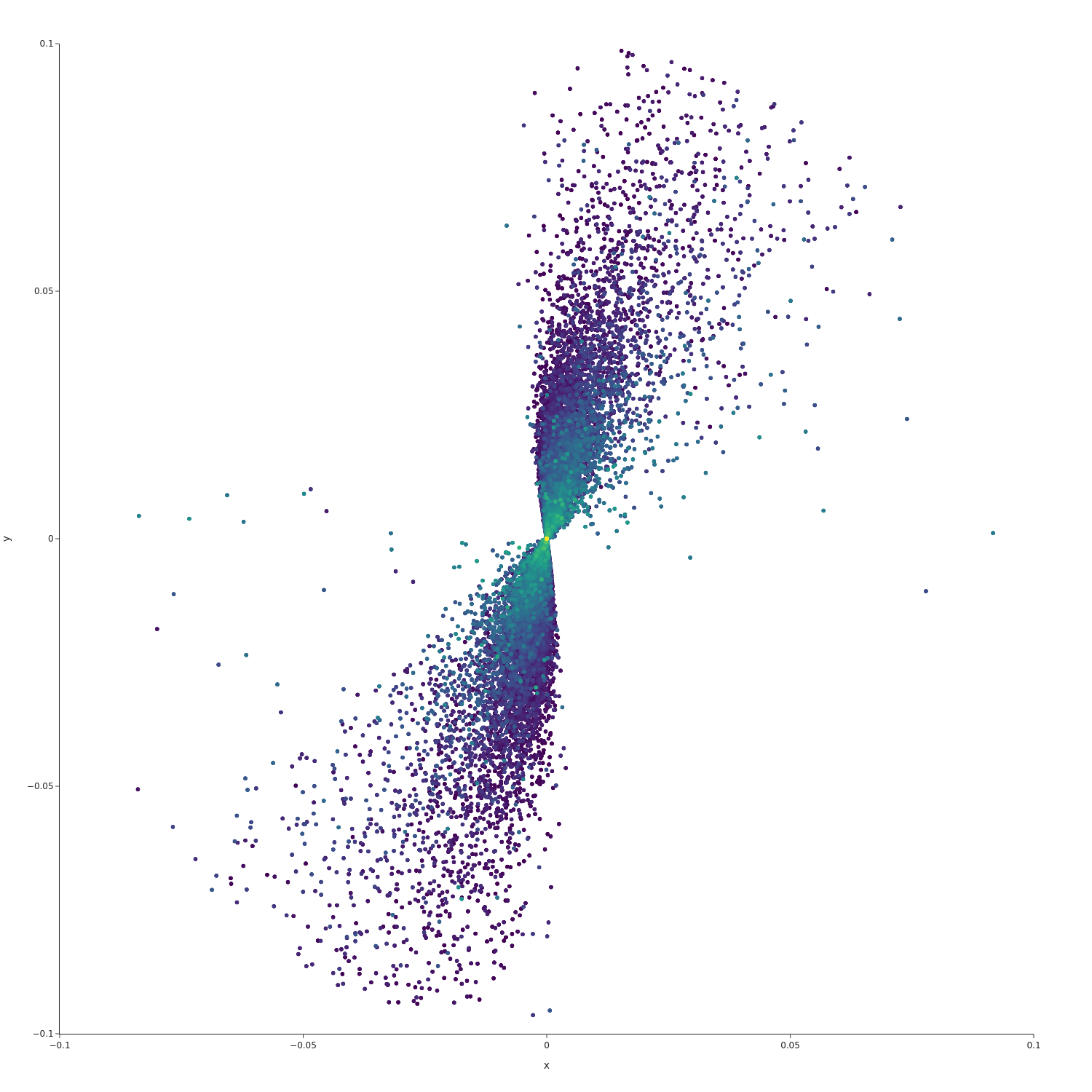}
            \caption{Original}
        \end{subfigure}
\hspace{0em}
        \begin{subfigure}[b]{0.49\textwidth}
            \centering
            \includegraphics[width=\textwidth]{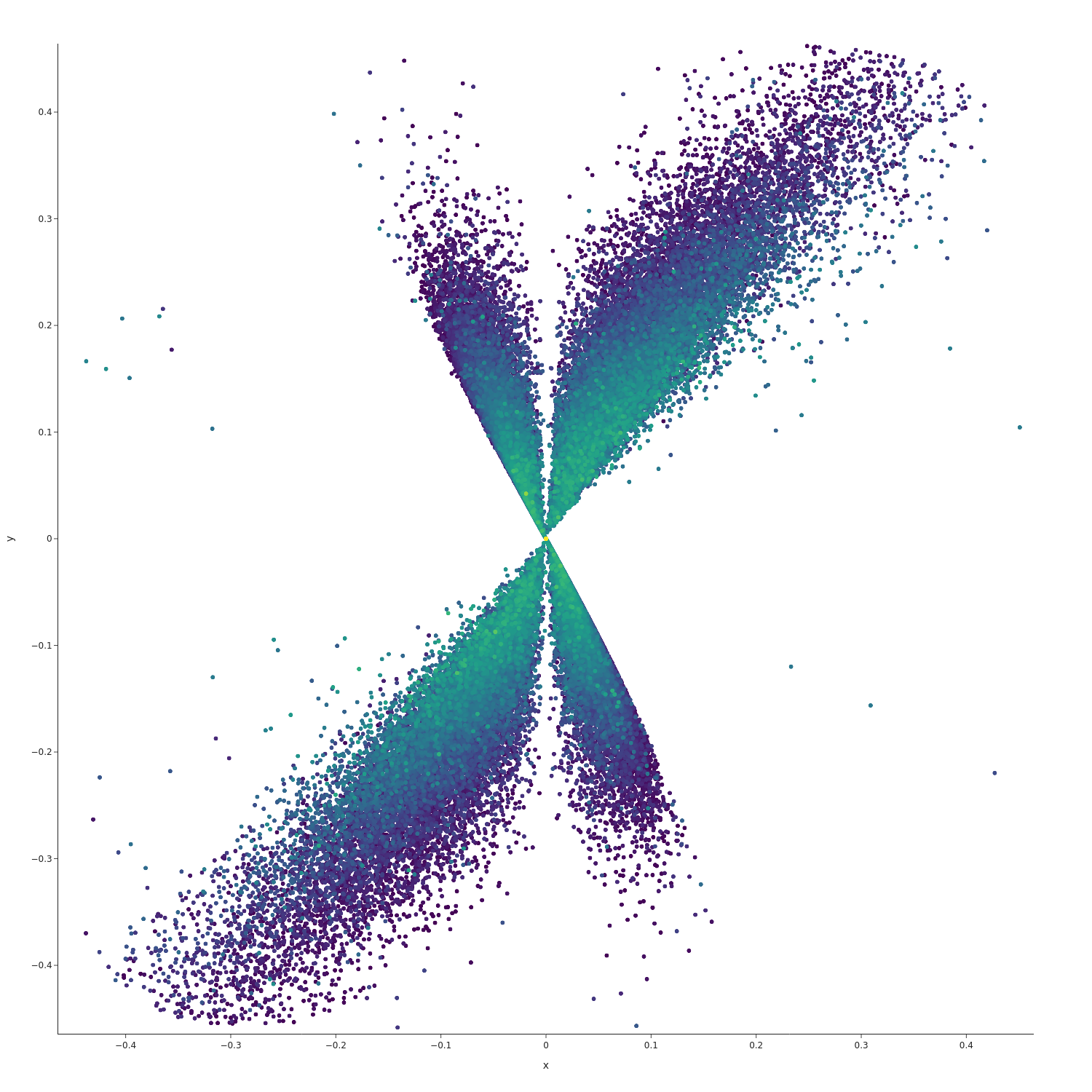}
            \caption{After pre-processing}
        \end{subfigure}

        \caption{\textbf{Example of correctly predicted clockwise helicity.} In this case, the original image (left) clearly shows a clockwise direction of rotation, since the extremities of the helix show that the majority of the points is found in the clockwise direction. The transformed image (right) shows the same pattern, but it does not have the same physical meaning anymore.}
        \label{fig:clockwise_correct}
    \end{figure*}

Clockwise helicity is visually characterized by a clockwise direction in rotation, as presented in Figure \ref{fig:clockwise_correct}. In this case, the helicity is rather clear, since the part of the helix in the clockwise direction has more sparse points than the counter-clockwise part. The same pattern can be recognized also in the transformed image, even though the effect of the application of the cubic root is that the image does not have the same physical meaning as before. In this case, the helicity direction was predicted correctly by the two classes classifier.

\begin{figure*}[!ht]
        \begin{subfigure}[b]{0.49\textwidth}
            \centering
            \includegraphics[width=\textwidth]{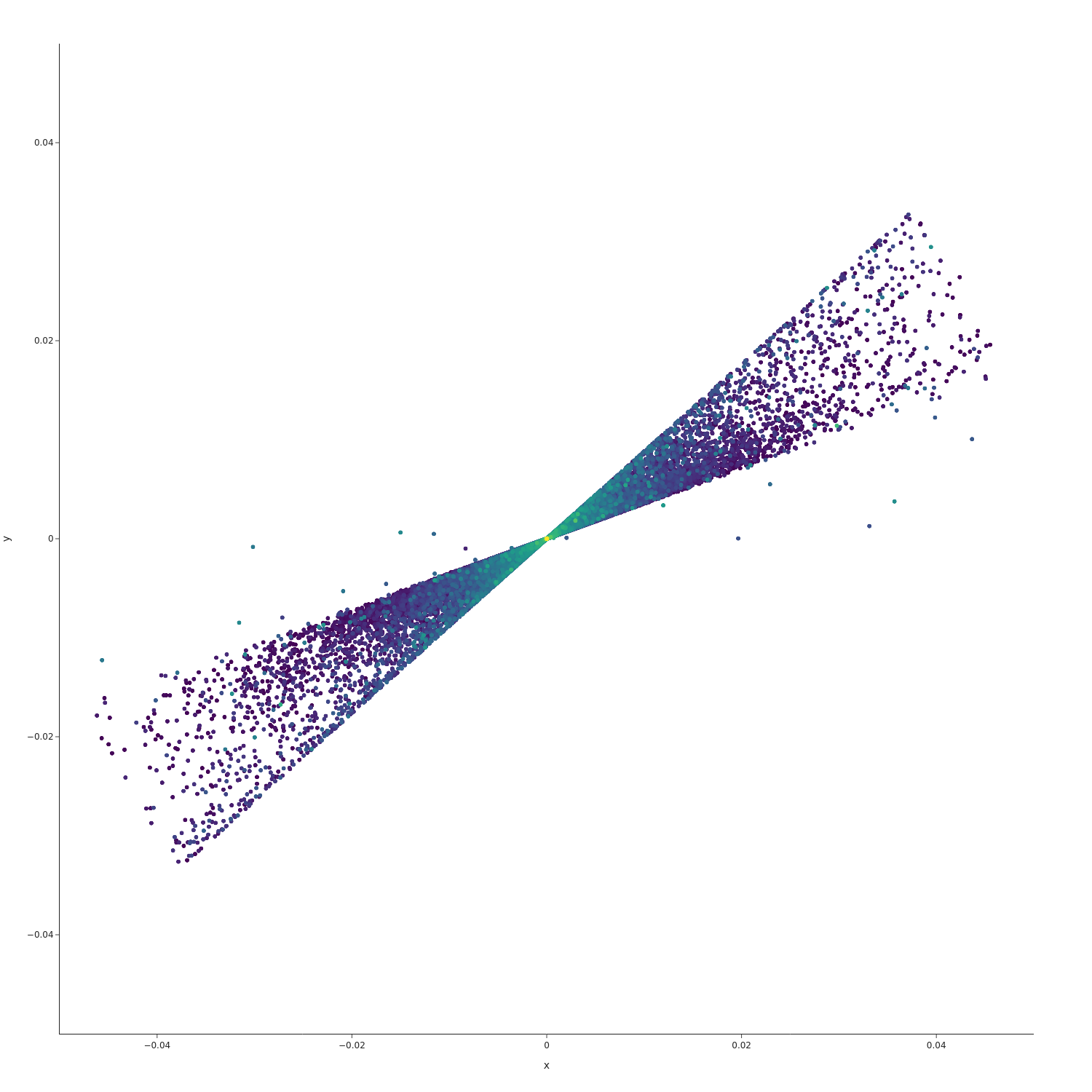}
            \caption{Original}
        \end{subfigure}
\hspace{0em}
        \begin{subfigure}[b]{0.49\textwidth}
            \centering
            \includegraphics[width=\textwidth]{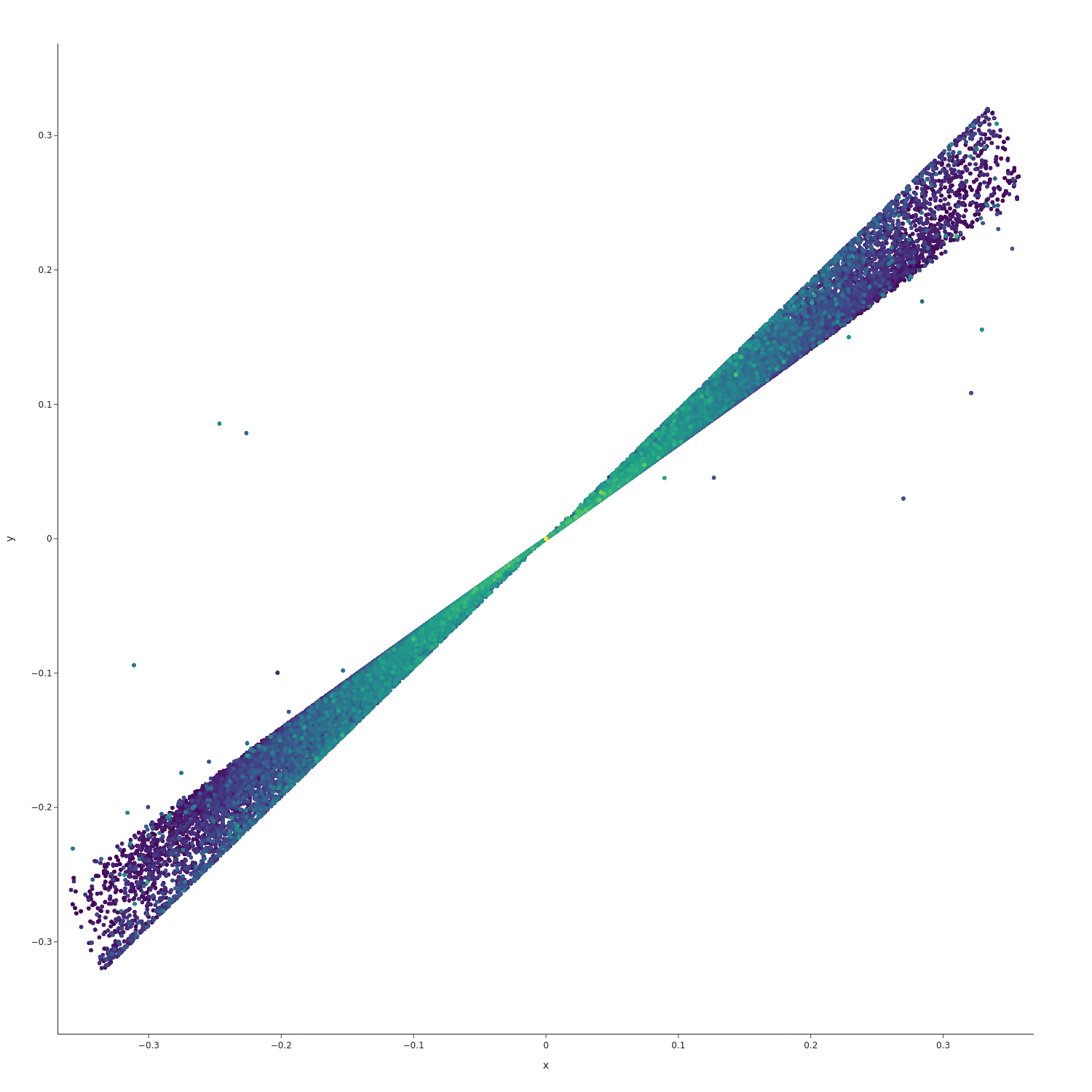}
            \caption{After pre-processing}
        \end{subfigure}

        \caption{\textbf{Example of incorrectly predicted clockwise helicity.} In this case, the original image (left) does not look clearly defined. The transformed image (right) is not clearly defined as well, but it emphasizes the presence of more sparse points in the counter-clockwise direction and the presence of a more continuous line in the other direction. This feature may be the reason why this clockwise helix is classified as a counter-clockwise helix.}
        \label{fig:clockwise_not}
    \end{figure*}

On the other hand, Figure \ref{fig:clockwise_not} presents an example of a clockwise helix that has been incorrectly classified as counter-clockwise by the two-classes classifier. In this case, the visual difference is more difficult to observe. The main difference is the presence of a more dense line in the counter-clockwise direction than in the clockwise direction, which is more challenging to be detected.

\subsubsection{Counter-clockwise helicity}
    
\begin{figure*}[!ht]
        \begin{subfigure}[b]{0.49\textwidth}
            \centering
            \includegraphics[width=\textwidth]{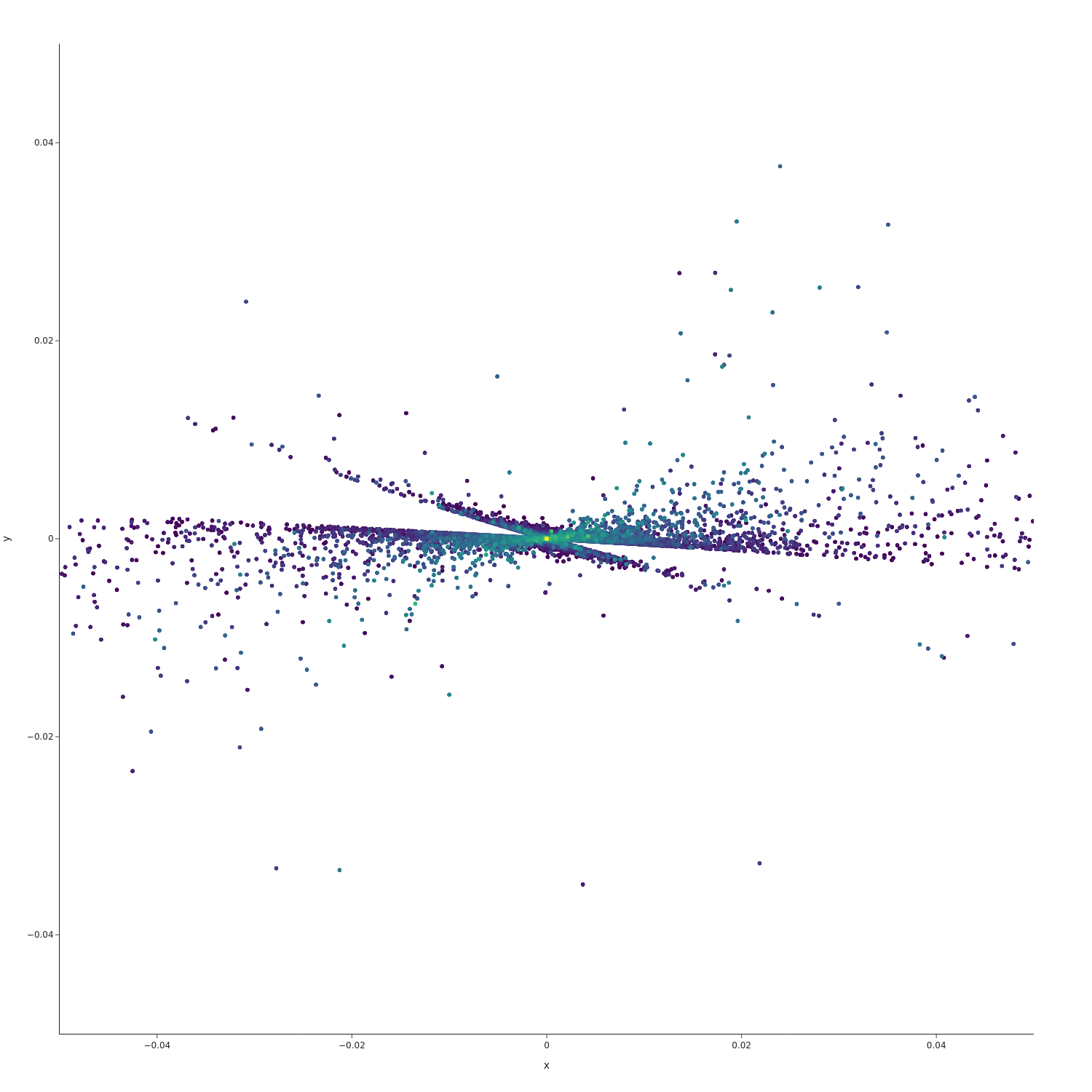}
            \caption{Original}
        \end{subfigure}
\hspace{0em}
        \begin{subfigure}[b]{0.49\textwidth}
            \centering
            \includegraphics[width=\textwidth]{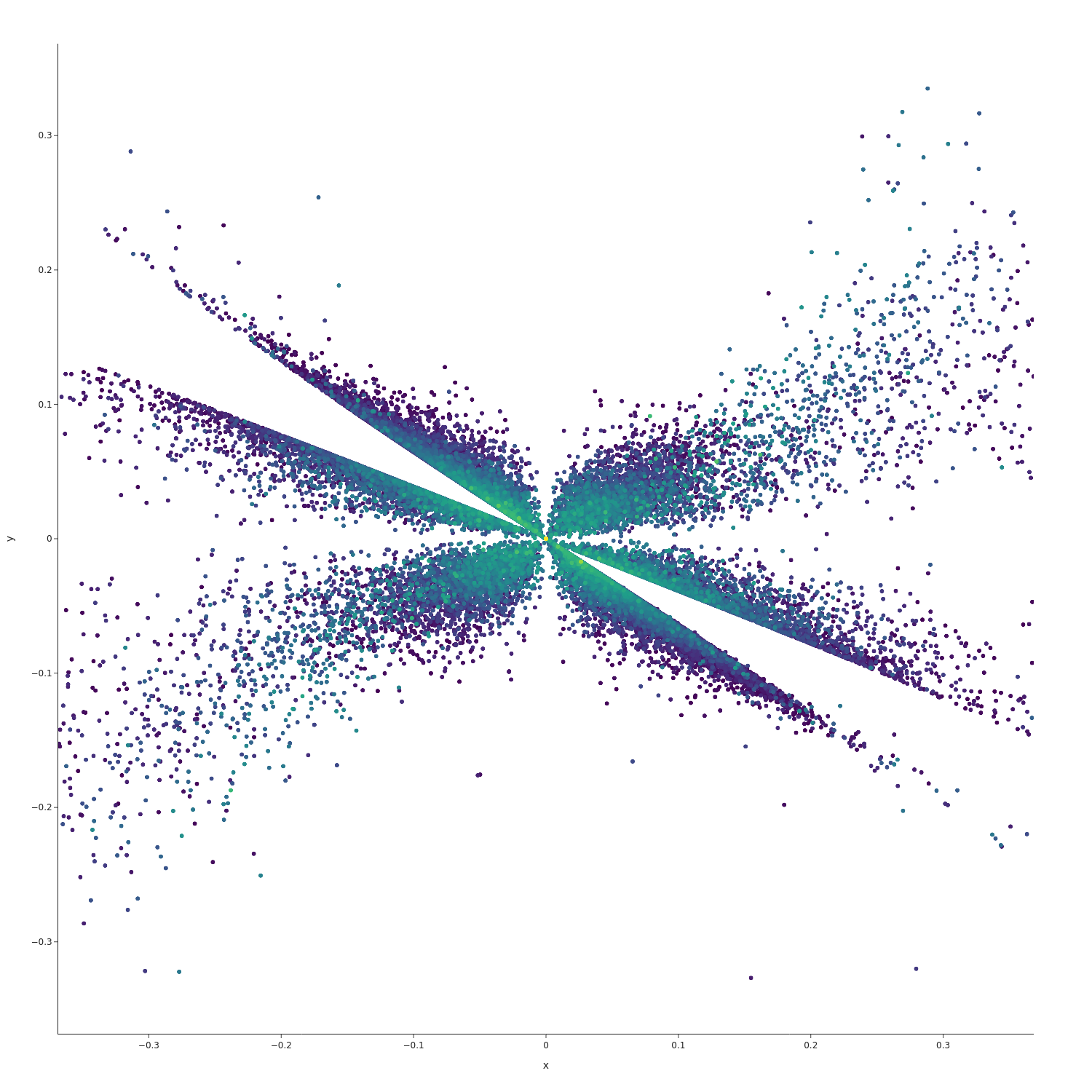}
            \caption{After pre-processing}
        \end{subfigure}

        \caption{\textbf{Example of correctly predicted counter-clockwise helicity.} In this case, the original image (left) suggests a counter-clockwise direction of rotation, since the extremities of the helix show that the majority of the points is found in the counter-clockwise direction. The transformed image (right) shows a similar pattern, but the transformation creates additional parts of the image, which are not related to the original physical meaning.}
        \label{fig:counterclockwise_correct}
    \end{figure*}

An example for counter-clockwise helicity which is classified correctly is presented in Figure~\ref{fig:counterclockwise_correct}. 

\begin{figure*}[!ht]
        \begin{subfigure}[b]{0.49\textwidth}
            \centering
            \includegraphics[width=\textwidth]{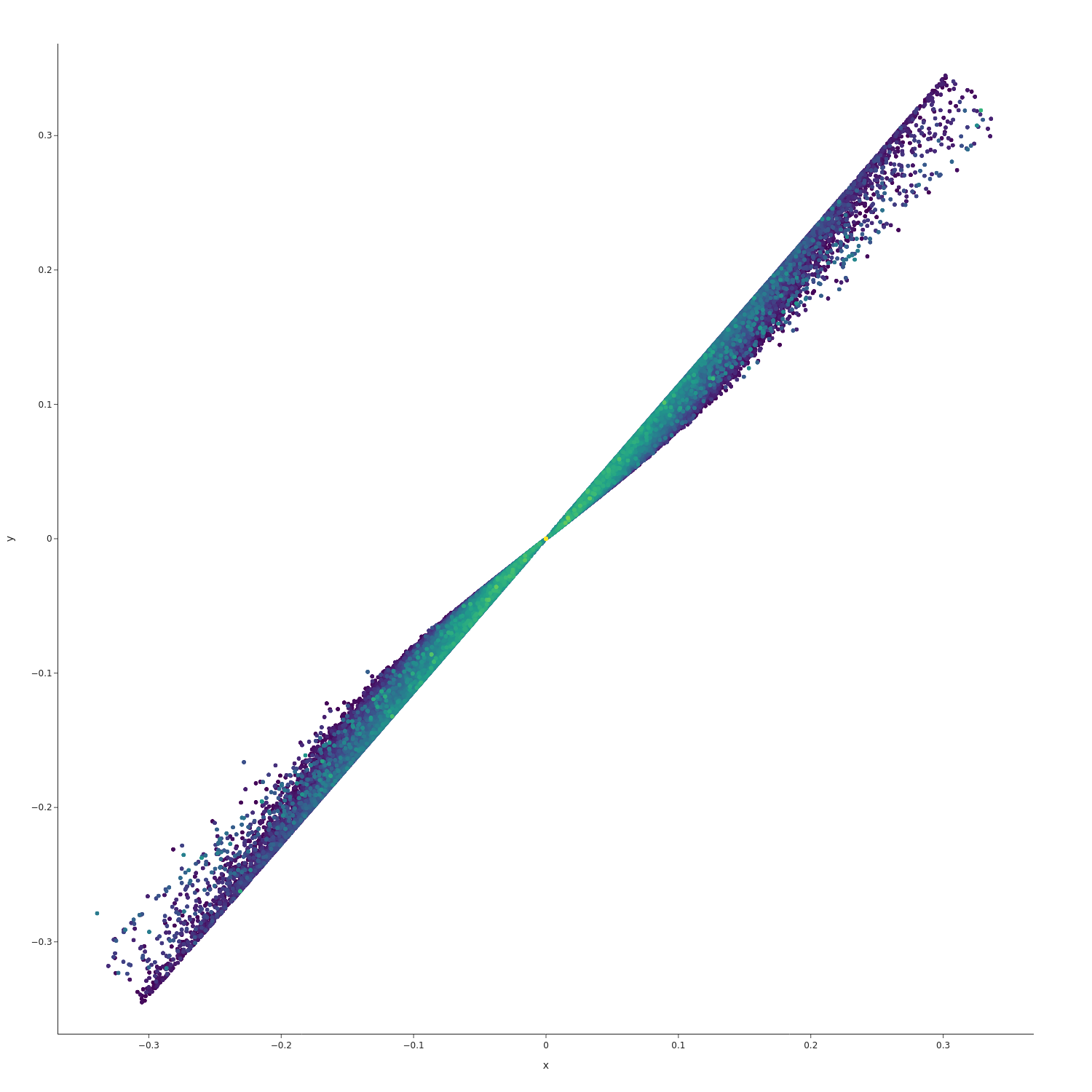}
            \caption{Original}
        \end{subfigure}
\hspace{0em}
        \begin{subfigure}[b]{0.49\textwidth}
            \centering
            \includegraphics[width=\textwidth]{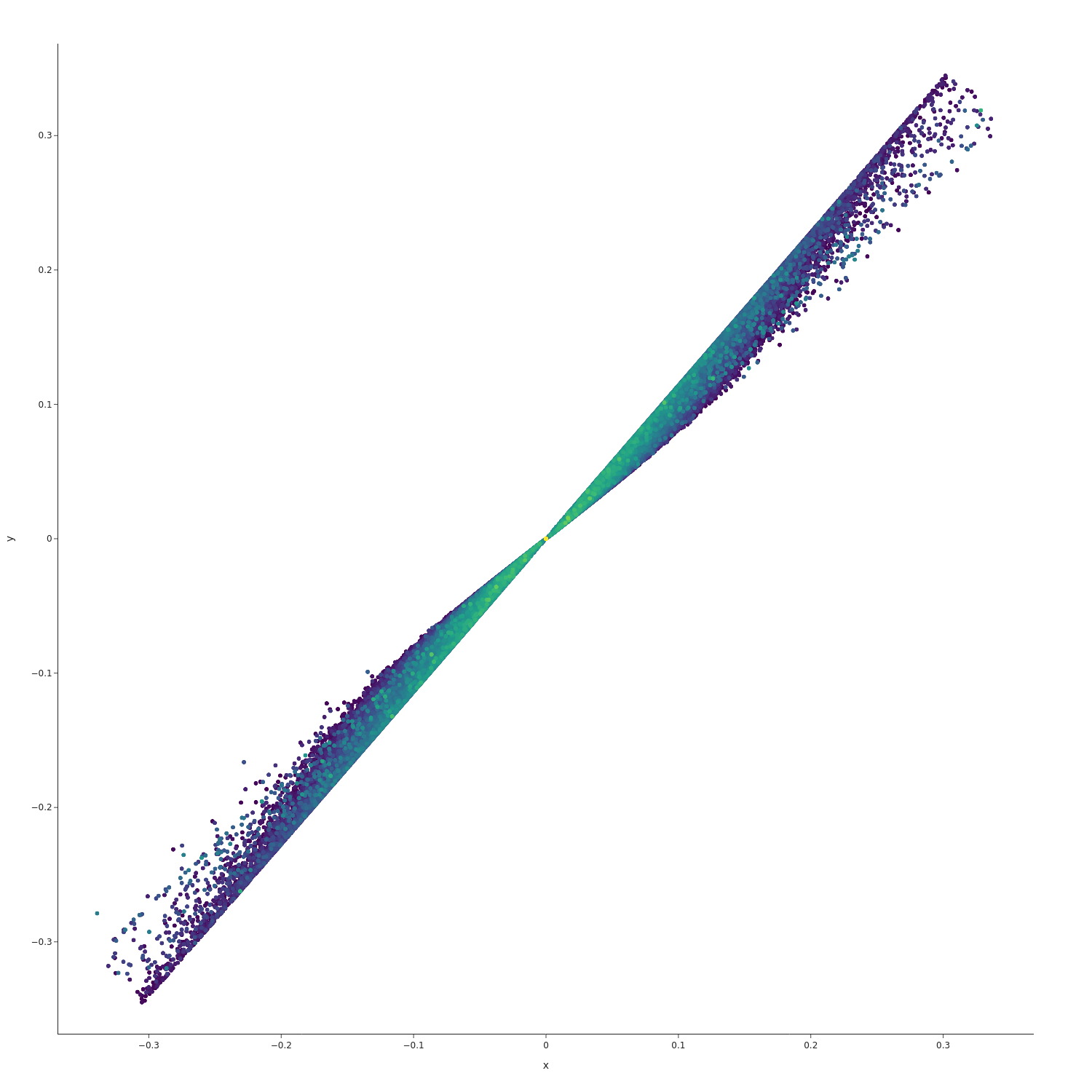}
            \caption{After pre-processing}
        \end{subfigure}

        \caption{\textbf{Example of incorrectly predicted counter-clockwise helicity}. In this case, the classifier predicted a counter-clockwise helicity, but this is a clockwise helix. In this case, the confusion may have been caused by the fact that this helix is almost flat and, therefore, more similar to a counter-clockwise helix than other samples.}
        \label{fig:counterclockwise_not}
    \end{figure*}

On the other hand, Figure \ref{fig:counterclockwise_not} shows an example of incorrect classification of the two classes classifier, which predicts a clockwise helicity. In this case, the helix looks rotated clockwise from a visual point of view, even though the generator parameter was set to counter-clockwise helicity. This resemblance may be the origin of the wrong classification.

\subsubsection{Zero helicity}

\begin{figure*}[!ht]
        \begin{subfigure}[b]{0.49\textwidth}
            \centering
            \includegraphics[width=\textwidth]{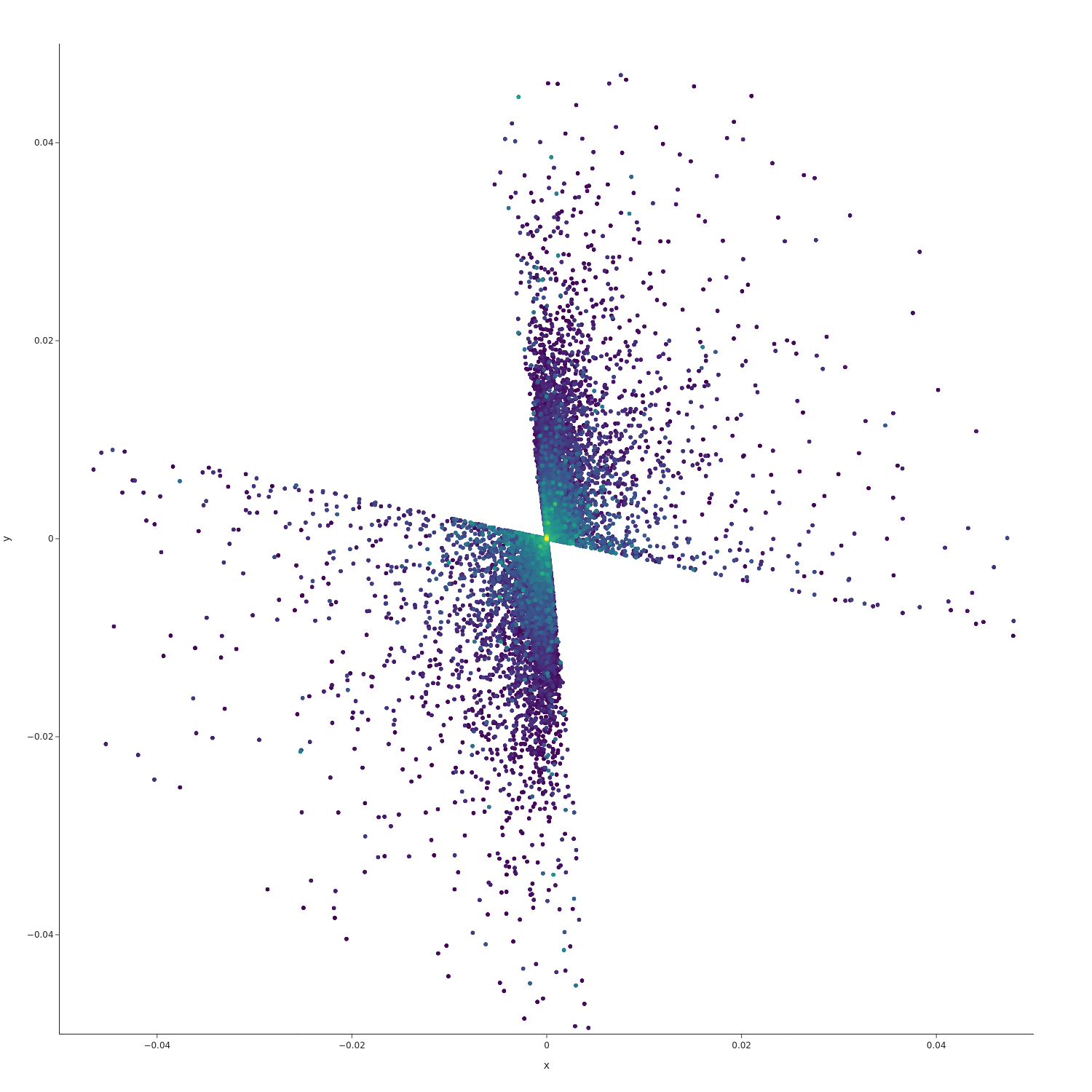}
            \caption{Original}
        \end{subfigure}
\hspace{0em}
        \begin{subfigure}[b]{0.49\textwidth}
            \centering
            \includegraphics[width=\textwidth]{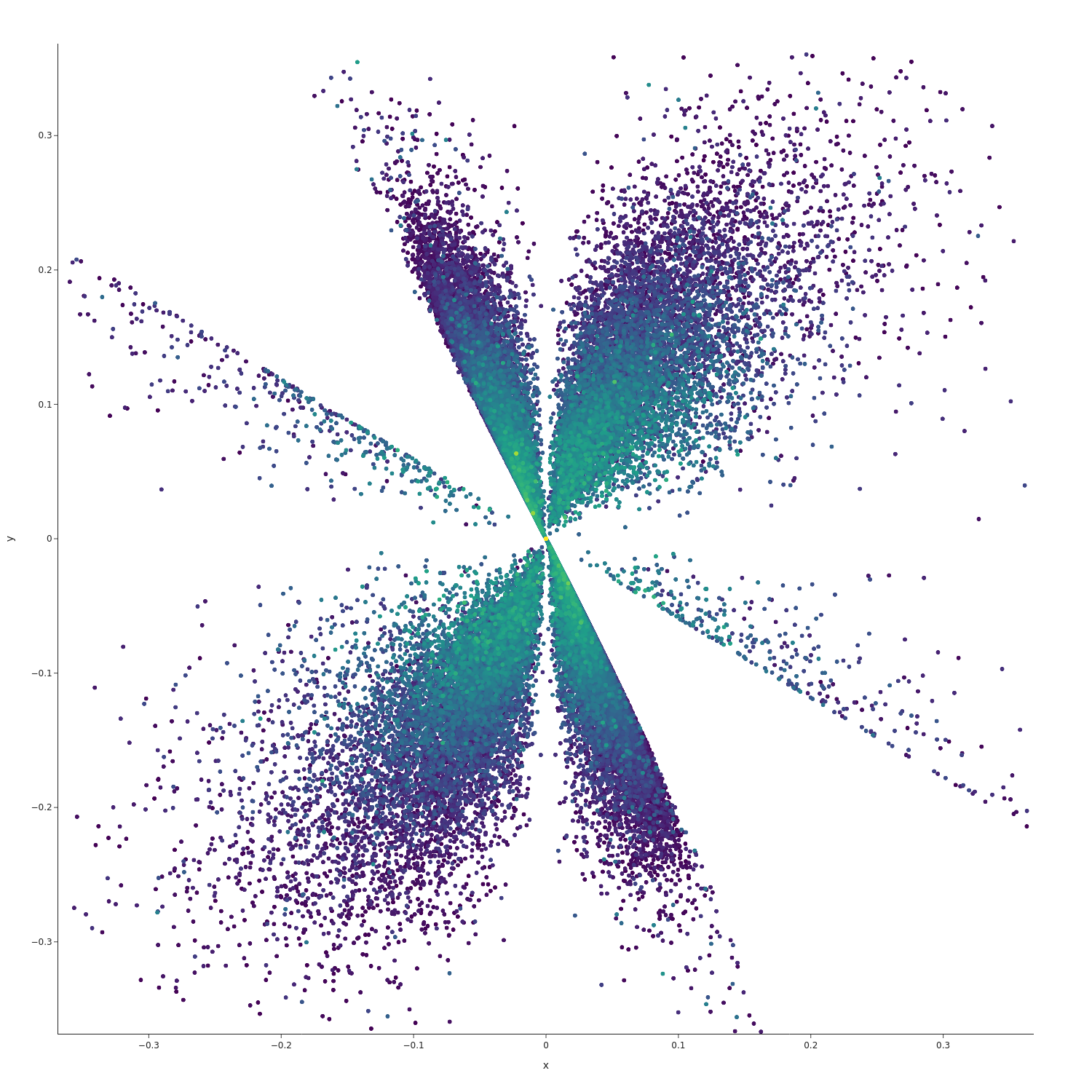}
            \caption{After pre-processing}
        \end{subfigure}

        \caption{\textbf{Example of correctly predicted zero helicity.} In this case, the clouds of points for both the original and the transformed plot are not evidently directed towards the clockwise or the counter-clockwise direction. This can be the reason why zero helicity was predicted correctly.}
        \label{fig:zero_correct}
    \end{figure*}

This case is particularly challenging since zero helicity is an intermediate case between clockwise and counter-clockwise helicity. Some examples, for this reason, are visually similar to the other two cases. In principle, the zero helicity cases should be represented as lines, but this is often not the case.
Figure~\ref{fig:zero_correct} presents a graph that has been correctly classified by the three classes classifier. What determines the zero helicity, in this case, is the absence of a clear helix direction, different from the previous examples. In the transformed version, it is possible to observe the presence of clouds of points that spread in both the clockwise and the counter-clockwise direction.
    
\begin{figure*}[!ht]
        \begin{subfigure}[b]{0.49\textwidth}
            \centering
            \includegraphics[width=\textwidth]{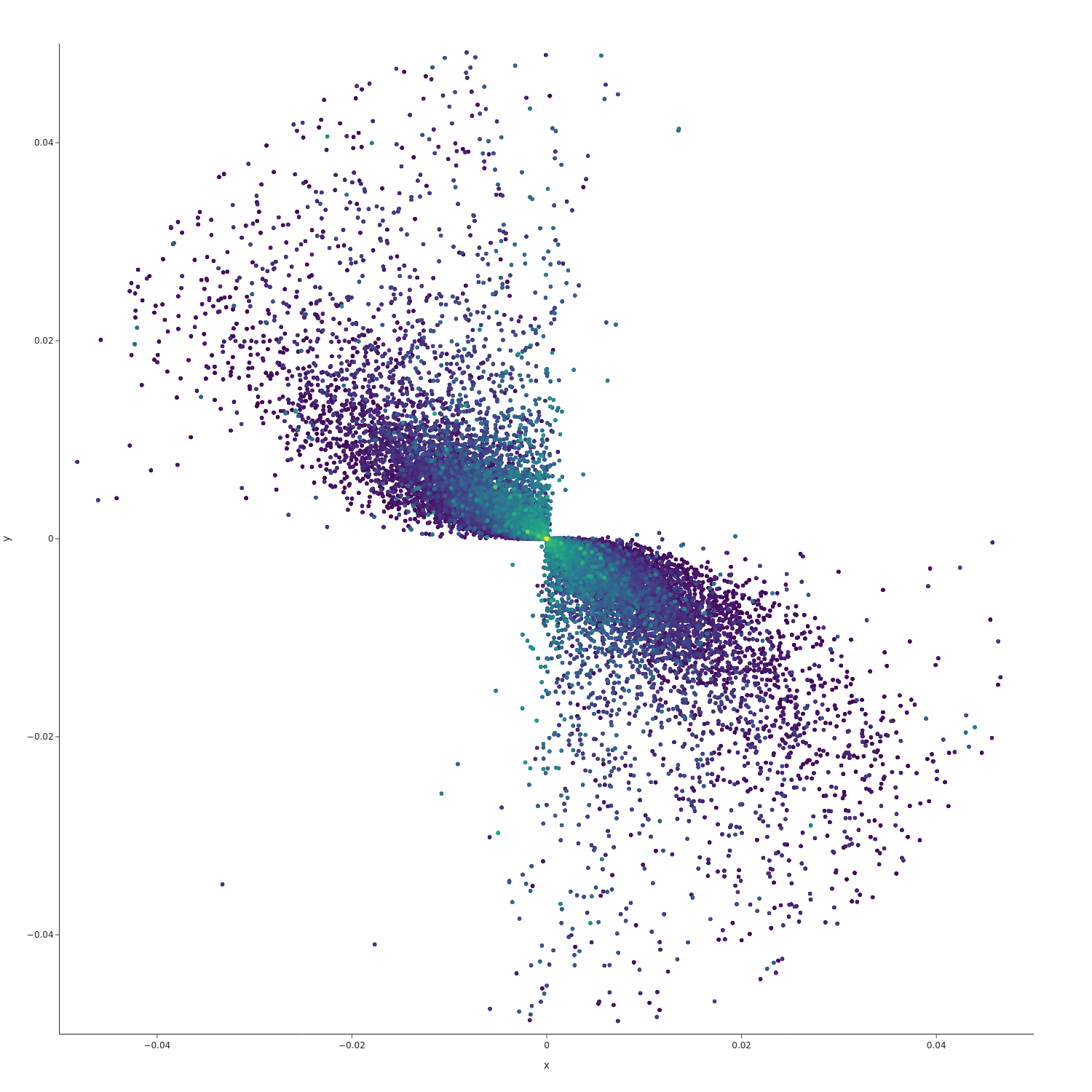}
            \caption{Original}
        \end{subfigure}
\hspace{0em}
        \begin{subfigure}[b]{0.49\textwidth}
            \centering
            \includegraphics[width=\textwidth]{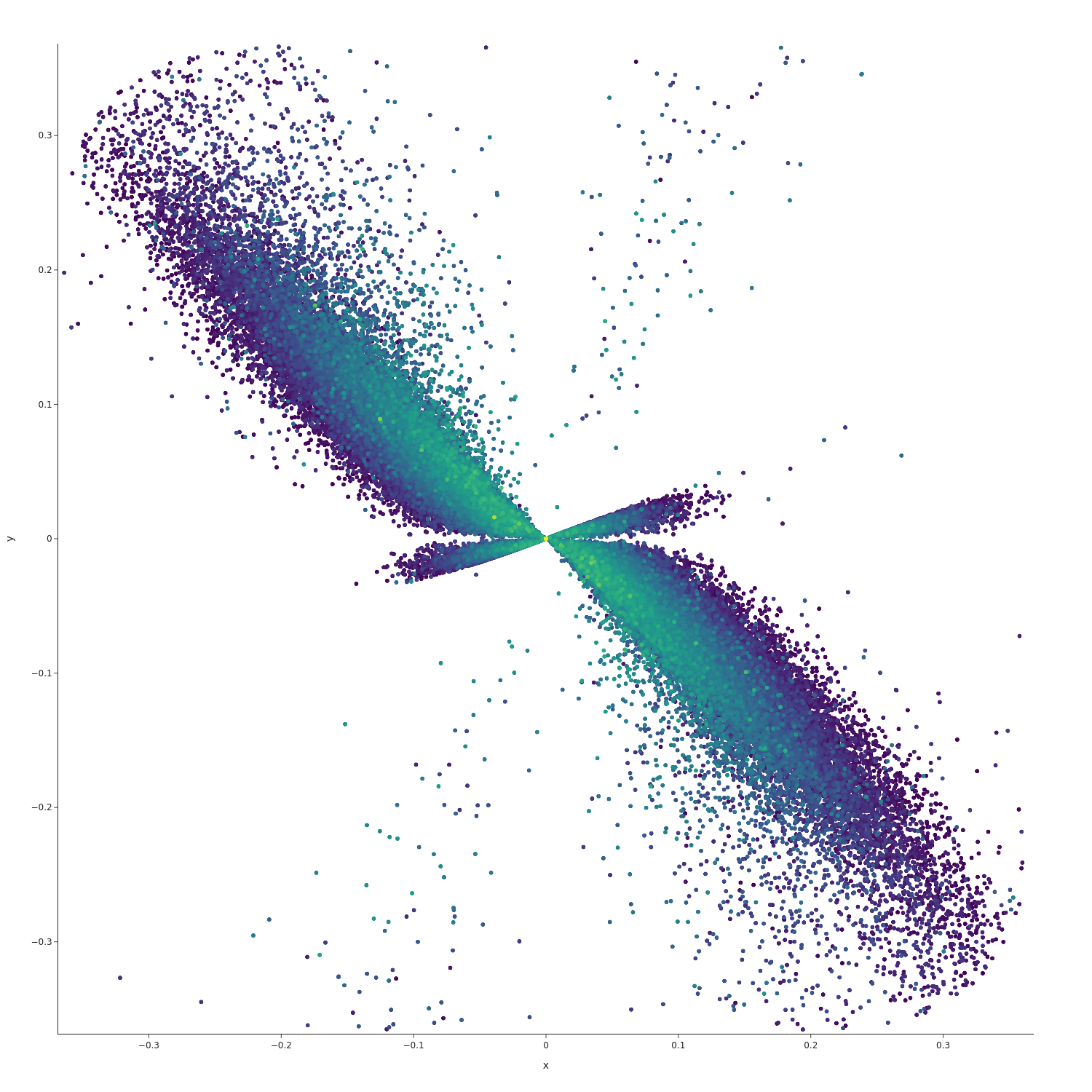}
            \caption{After pre-processing}
        \end{subfigure}

        \caption{\textbf{Example of incorrectly predicted zero helicity.} In this case, the three classes classifier classifies the plot as representing a clockwise helix. A possible reason why this happens is that the cloud of points spreads slightly more in the clockwise direction.}
        \label{fig:zero_not}
    \end{figure*}

On the other hand, Figure \ref{fig:zero_not} presents a more ambiguous case, and the plot resembles the one of clockwise helicity since the points are more spread in the clockwise direction. This sample's helicity, indeed, is classified as clockwise by the three classes classifier. The transformed plot, in this case, shows the same pattern as the original image, with more points in the clockwise direction.

\end{document}